\documentstyle[]{mn}

\input{psfig.sty}

\ifnfsstwo

\fi

\ifnfssone
  \newmathalphabet{\mathit}
    \addtoversion{normal}{\mathit}{cmr}{m}{it}
    \addtoversion{bold}{\mathit}{cmr}{bx}{it}

\fi

\ifoldfss

\fi

\loadboldmathitalic
\loadboldgreek

\def\Zsun{\thinspace\hbox{$\hbox{Z}_{\odot}$}}
\def\msun{\thinspace\hbox{$\hbox{M}_{\odot}$}}

\def\gal{galaxy}
\def\gals{galaxies}
\def\el{elliptical}
\def\sph{spheroid}

\def\ev{evolution}

\def\for{formation}
\def\sfor{star formation}

\def\cf{cooling flow}
\def\gw{galactic wind}
\def\lum{luminosity}
\def\halpha{\mbox{H$\alpha$}}
\def\hbeta{\mbox{H$\beta$}}
\def\lya{Lyman~$\alpha$}

\def\lbgs{Lyman break galaxies}
\def\C2{DSF 2237+116 C2}
\def\kms{\,km~s$^{-1}$}
\def\gr{$G-{\cal R}$}
\def\grc{$(G-{\cal R})_{calc}$}
\def\gro{$(G-{\cal R})_{obs}$}

\title[Lyman break galaxies and spheroids]
       { Lyman break galaxies as young spheroids}
\author[A. C. S. Fria\c ca and R. J. Terlevich]
       {A. C. S. Fria\c ca$^1$ and R. J.
Terlevich$^{2,3}$\\
$^1$Instituto Astron\^omico e Geof\'\i sico, USP,
Caixa Postal 3386, 01065-970 S\~ao Paulo, SP, Brazil\\
$^2$Institute of Astronomy,
Madingley Road, Cambridge CB3 0EZ, UK\\
$^3$Visiting Professor at Instituto Nacional de Astrof\'{\i}sica,    
Optica y Electr\'{o}nica. Av. Luis Enrique Erro 1, Tonanzintla, Puebla,
Mexico}

\pubyear{1998}
\begin{document}

\maketitle

\begin{abstract}
We investigate  the nature of \lbgs\ (LBGs) using
a chemodynamical model for \ev\ of  \gals.
Our models predict an early (the first Gyr) stage of intense \sfor\  in
the \ev\ of massive \sph s which could be identified to
the LBGs, observed at redshift $\sim 3$ with strong ongoing \sfor.
In particular,
we are successful in reproducing the properties of  the LBG \C2\ with
a model describing a young $\sim L^*$ \sph.
The comparison of the predictions of our models with the observations
gives support to the scenario in which LBGs are the progenitors
of present-day massive spheroids, i.e. bulges of luminous early type
spirals or 
luminous elliptical galaxies.

\end{abstract}


\section{Introduction}

Colour selection techniques based on 
the Lyman limit break of the spectral energy distribution
caused by neutral hydrogen absorption
have been used for many years in surveys for distant 
QSOs (e.g. Warren et al.\ 1987).  
Guhathakurta et al.\ (1990) and Songaila, Cowie \& Lilly (1990) 
used this method to set limits on 
the number of star-forming galaxies at $z \approx 3$ in faint galaxy 
samples.
More recently,  Steidel \& Hamilton (1992, 1993) and Steidel, Hamilton \&
Pettini 
(1995), using this method, 
designed a broad band filter set (the $U_n G {\cal R}$ system), which
allowed them
to discover a widespread population of 
star forming galaxies at redshift $z \simeq 3$,
the Lyman break \gals\ (LBGs).
Spectroscopic confirmation
of their redshifts was first presented by Steidel et al. (1996),
and WFPC2 images of select LBGs were published by 
Giavalisco, Steidel \& Macchetto (1996).

An important recent advance in the study of LBGs was the
availability of the first results from a program of near-infrared
spectroscopy
aimed at studying the familiar rest-frame optical 
emission lines from H~II regions of LBGs (Pettini et al 1998b, hereafter
P98). 
The program was successful in detecting 
Balmer and [O~III] emission lines in five LBGs.
The nebular luminosities imply
star formation rates (SFRs)  larger than those deduced from the UV
continuum,
which suggests significant dust reddening.
In four LBGs the velocity dispersion of the emission lines is 
$\sigma_{em} \simeq 70$~\kms, while the fifth system has $\sigma_{em}
\simeq 200$~\kms.
The relative redshifts of 
interstellar absorption, nebular emission, and \lya\ emission lines
differ by several hundred \kms\ , a similar
effect to that found in nearby HII galaxies ( Kunth et al 1998) indicating
that large-scale 
outflows may be a common characteristic of both starbursts and LBGs.

On the other hand, we have developed a chemodynamical model
(Fria\c ca \& Terlevich 1994; Fria\c ca \& Terlevich 1998, hereafter FT)
for \for\ and \ev\ of
\sph s, which are suspect to be the $z=0$ counterparts of LBGs (Steidel et
al. 1996).
Our chemodynamical model
combines multi-zone chemical evolution with 1-D hydrodynamics
to follow in detail the evolution and radial behaviour 
of gas and stars during the formation of an \sph.
The star formation and the subsequent stellar feedback regulate
episodes of wind, outflow, and cooling flow.
The  knowledge of the radial gas flows in the galaxy allows us to
trace metallicity gradients,
and, in particular, the formation of a high-metallicity core in \el s.
The first $\sim 1$ Gyr of our model \gals\ shows
striking similarities to the LBGs:
intense star formation, compact morphology, 
the presence of  outflows, and significant metal content.
We now proceed to examine these similarities, and, in particular,
to consider the implications
of the recent near-infrared observations of  P98.
We demonstrate that our model supports the scenario in which
LBGs are the progenitors of  the present-day bright spheroids.
In this paper, the SFRs, luminosities and sizes quoted
by P98 are converted to the cosmology adopted here
($H_0 = 50$~km~s$^{-1}$~Mpc$^{-1}$, $q_0 = 0.5$).

\section{Lyman break galaxies as young spheroids}

There are several evidences in favour of the LBGs being the high-redshift
counterparts of the present-day spheroidal component of 
luminous \gals\ (Steidel et al. 1996, Giavalisco et al. 1996):
their comoving space density is at least 25 \% of that of
luminous ($L \geq L^*$) present-day \gals;
the widths of the UV interstellar absorption lines
in their spectra imply velocity dispersions of $180-320$ km s$^{-1}$,
typical of the potential well depth of luminous \sph s;
they have enough binding energy to remain relatively compact despite the
very
high SN rate implied by their SFRs.
In addition, the population of LBGs shows strong clustering
in concentrations which may be the precursors of the present rich
clusters of \gals\ at a time when they were
beginning to decouple from the Hubble flow
(Steidel et al. 1998).
In the context of Cold Dark Matter models of structure formation,
the LBGs must be associated with very large halos, of mass $\ga 10^{12}$
\msun, in order to have developed such strong clustering at $z \sim 3$.

Assuming a Salpeter IMF, 
P98 inferred from the emission Balmer lines 
values for the SFR (uncorrected for dust)  of their LBGs in the range
$19-210 h_{50}^{-2}$ \msun\ yr$^{-1}$.
These values are typically a factor of several larger than
those deduced from the UV continuum and indicate that
the correction for dust is typically 1-2 magnitudes at 1500 \AA.
Dickinson  (1998), for a large sample of LBGs,
deduced from the UV continuum SFRs in the range
$ 3-60 h_{50}^{-2}$ \msun\ yr$^{-1}$.
Assuming a 1 Gyr old continuous \sfor, he used the \gr\ colours
to compute corrections for dust extinction to the SFR.
With a Calzetti (1997) attenuation law, 
after correction for dust extinction,
the SFR range becomes $\sim3 - \sim 1500$ \msun\ yr$^{-1}$.
These levels of \sfor\ are remarkably close to 
the values of the SFR  exhibited in the early \ev\ of
the chemodynamical models of FT.

FT built a sequence of  chemodynamical models reproducing the main
properties
of elliptical galaxies.
The calculations begin with a gaseous protogalaxy
with initial baryonic mass $M_G$.
Intense star formation during the early stages of the \gal\  builds up
the stellar body of the \gal,
and during the \ev\ of the \gal, gas and stars exchange mass
through star formation and stellar gas return.
Owing to inflow and \gw\ episodes occuring during the \gal\ \ev,
its present  stellar mass is $\sim15-70$\% higher than $M_G$.
Gas and stars are embedded in a dark halo of core radius $r_h$
and mass $M_h$ (we set $M_h=3M_G$).
The models are characterised by $M_G$, $r_h$, and a \sfor\ prescription.
The SFR is given by a Schmidt law $\nu_{SF} \propto \rho^{n_{SF}}$ 
($\rho$ is the gas density and 
$\nu_{SF} = SFR/\rho$ is the specific SFR).
Here we consider the standard \sfor\ prescription of FT, 
in which the normalization of $\nu$ is $\nu_0=10$ Gyr$^{-1}$
(in order to reproduce the suprasolar [Mg/Fe] ratio of giant \el s),
$n_{SF}=1/2$,
and the stars form in a Salpeter IMF from 0.1 to 100 \msun.
A more detailed account of the models can be found in FT.
Figure 1 shows the \ev\ of the SFR
for models with $M_G$ in the range 
$5\times 10^9 - 5\times 10^{11}$ \msun\ ($r_h=0.8-5$ kpc).
During the maximum of  the SFR, the stellar velocities dispersions of
these models,
$55-220$ \kms, bracket the $\sigma_{em}=55 - 190$ \kms\ range
of the P98's LBGs.
The corresponding present-day (age of 13 Gyr) luminosities
are $0.05 L^* - 1.4 L^*$ ($-M_B=17.6-21.3$).
For our models, the typical range of SFR averaged over the first Gyr, 
$10-700$ \msun\ yr$^{-1}$,
reproduces well the SFRs found for LBGs, deduced from
both the Balmer lines and the UV continuum corrected for dust extinction.
In addition, the SFR drops dramatically after 1.5-2 Gyr, and becomes
below the lowest SFRs found for the LBGs.
The similarity of of the SFRs of our models to those of LBGs
allows us to identify the LBGs to young ($ \la 1-2$ Gyr) spheroids.

It is important to note that
the moderately high SFRs of the LBGs
seem to be difficult to conciliate with the predictions of
the simplistic one-zone (or monolithic) models of 
formation of \el\ \gals\ for supra-$L^*$ systems.
The monolithic models of \for\ of early-type \gals\
have been worked out in the early 1970's 
(e.g. Larson 1975) and are succesful at reproducing 
the supra-solar [Mg/Fe] of  bright \el s
(Matteuccci \& Tornamb\'e 1987; Hamann \& Ferland 1993),
but the required short \sfor\ time scale ($\sim10^8$ yr)
implies extremely high SFRs during the formation of $L>L^*$ \el s.
As a matter of fact, in the one-zone model,
a gaseous protogalaxy with $5\times 10^{10}$ \msun,
would have a peak SFR of $\sim 5000$ \msun yr$^{-1}$,
and a present-day $M_B=-21.1$.
At least at redshift $3 \la z \la 3.5$,
such  SFR is excluded by 
the properties of  the population of  LBGs.
By contrast, 
in the chemodynamical model, 
the metallicity and abundance ratios of the central
region of the young elliptical are explained with no need for all
the galaxy having a global starburst coordinated with the central
starburst,
which avoids the excessively high SFRs of the one-zone model.
The most massive model  here
($M_G=5\times 10^{11}$ \msun; present-day $M_B=-21.3$)
has a peak SFR of 1050 \msun yr$^{-1}$,
consistent, after correction for dust extinction, 
with the highest SFRs derived from the UV continuum of LBGs (Dickinson
1998).
Note that, as we show below,
because the observed rest-frame UV colours
limit the amount of dust extinction to $\sim 3$ mag at most,
we cannot evoke dust to hide a 5000 \msun\ yr$^{-1}$ starburst
as a LBG at $z \sim 3$.

HST optical imaging, which probes the rest frame UV between 1400 and 1900
\AA,
has revealed that the LBGs are generally compact,
with  a typical half-light radius
of $1.4-2.1 h_{50}^{-1}$ kpc  (Giavalisco et al. 1996).
The observed LBGs do not seem to have disk morphology,
with the exception of a few objects without central concentration.
In addition, some objects have a light profile  following a $r^{1/4}$ law 
over a large radial range,
which supports the identification of LBGs to young spheroids.
Near infrared imaging have yielded half-light radii
in the range $1.7-2.3 h_{50}^{-1}$ kpc (P98).
The similarity of the near-infrared sizes to those obtained
by the HST suggests that
the optical morphology follows the UV morphology.
As shown in the next section, the compact appearance of
the LBGs, both in the UV and in the optical, is reproduced
by our young spheroid models.

Note that,
due to the strong fading of surface brightness with redshift 
($\propto (1+z)^{-4}$),
the outer parts ($r \ga 10$ kpc) of the \gal\ with milder
\sfor\ rates ($\nu_{SF} \sim 1$ Gyr$^{-1}$ or less) 
would be missed in high redshift observations.
The difficulty in observing the outer regions of the \gal\ would only be
compounded
if there is some dust extinction.
There is an analogy between the LBGs and nearby HII galaxies, 
in which we are observing only the brightest  part of the \gal,
superposed on much more extended low surface brightness object,
when deeper expositions are made available (Telles \& Terlevich 1997;
Telles, Melnick \& Terlevich 1997).
Additional support to the LBG-starburst connection comes from the fact
that
the  LBGs in the P98 sample 
fall on the extrapolation to higher luminosities
of the correlation $L_{\rm {H}\beta}- \sigma$ found for local H~II
galaxies
by Melnick, Terlevich, \& Moles (1988)
(Terlevich 1998).

 \begin{figure}
 \centerline{
 \psfig{figure=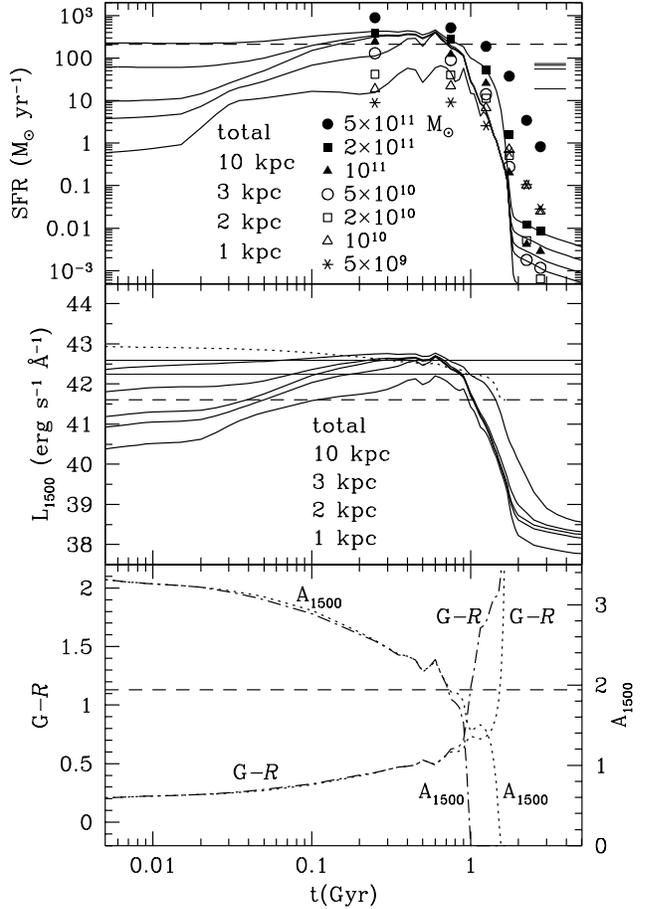,width=8.5cm,angle=0}}
 \caption{
Top panel: \ev\ of the SFR for the fiducial over the whole \gal\ and
inside several radii.
Also given the predicted SFRs averaged over the time spans
0-0.5, 0.5-1, 1-1.5, 1.5-2, 2-2.5, and 2.5-3 Gyr for several models
(symbols labelled by the model $M_G$).
Middle panel: evolution of the 1500 \AA\ luminosity over the whole \gal\
and
inside several projected radii.
The luminosities have been calculated from the models
of Bruzual \& Charlot (1998), for a Salpeter IMF from 0.1 to 100 \msun\
and
metallicities from $Z=0.0001$ to $Z=0.1$.
Lower panel: \ev\  of the unreddened  \gr\ colour for the fiducial model,
over the whole \gal\ and inside a projected radius of 10 kpc 
(dotted and dot-dashed lines, respectively).
Also shown (with the same line styles) the amount of reddening $A_{1500}$
needed for \gr\ of the model to match the observed \gr.
In all three panels
the dashed horizontal lines represent the respective observed quantities
for \C2:
SFR deduced from the \hbeta\ emission;
$L_{1500}$ uncorrected for dust;
\gr\ colour.
In the middle panel,
the two thin  horizontal lines denote the observed $L_{1500}$ 
with a correction for dust extinction deduced from the \gr\ colour
as in P98 (see text),
and the dotted line indicates the observed $L_{1500}$
corrected for dust extinction using the $A_{1500}$ value
over the whole \gal\ shown in the lower panel.
In the upper panel, the thin long dashes on the right denote
the SFRs deduced from \hbeta\ emission
for the $\sigma \approx 70$ \kms\ LBGs of P98;
the highest value corresponds in fact to an upper limit
for Q0000-263 D6, since for this object (at $z_{em}=2.966$),
\hbeta\ is outside the K-band, and the SFR was deduced from
the [O III]$\lambda 5007$ luminosity, 
assumimg \hbeta/[O III]$\lambda 5007 \leq 0.5$. 
}
\end{figure}

\section{\C2, a young $\sim L^*$  spheroid?}

It is of interest to compare the predictions for our models with
the observational data of  \C2, the most massive
LBG  (the LBG with the largest $\sigma_{em}$) in the P92 sample.
The properties of this object are successfully
described by the fiducial model of FT
($M_G=2\times 10^{11}$ \msun\ and $r_h=3.5$ kpc ).
Its present-day stellar mass, $2.4\times 10^{11}$ \msun,
corresponds  to $L_B = 0.7 L^*$,
which allows us to identify \C2\ to an $\sim L^*$ spheroid seen during
its early \ev, characterised by intense \sfor.
For the fiducial model, Figure 1 shows the \ev\ of the SFR  within several
radii.
The initial stage of violent \sfor\  lasts $\sim 1$ Gyr, and exhibits
a maximum SFR of $\sim 500$ $\msun$~yr$^{-1}$ at 0.6 Gyr.
After the \gw\ is established (at $t=1.17$ Gyr), the SFR plummets
and practically all \sfor\  within 10 kpc is concentrated
inside the inner kpc.
The late central \sfor, characterised by
a moderate SFR ($\sim$ few \msun yr$^{-1}$), 
is fed by a \cf\ towards the galactic centre.
The stagnation point separating the wind and the inflow moves inwards
until it reaches the galactic core at $t=1.8$ Gyr,
when a total wind is present throughout the galaxy.
After this time, indicating the end of the star-forming stage,
only very small levels of \sfor\ are present in the \gal.
The early stage of \sfor\, during which the stellar body of  the \gal\ is
formed
(the stellar mass reaches 50\% of its present value at $t=3.9\times 10^8$
yr),
resembles the LBGs.
The average SFR  during the first Gyr, 328 \msun\ yr$^{-1}$,
is very similar to the SFR of  210$h_{50}^{-2}$  \msun\ yr$^{-1}$
of \C2\ inferred from its \hbeta\ \lum.
In addition, the SFR is concentrated in the inner 2-3 kpc,
which gives to our model \gal\ the compact appearance typical of LBGs.

Figure 1 also shows $L_{1500}$,
the \lum\ at 1500 \AA, which allows a more direct comparison
with the imaging data.
Note that our models reproduce the compact appearance of LBG,
the light being concentrated in the inner $\sim 3$ kpc until the maximum
of the SFR and in the inner $\sim 2$ kpc after that time.
The luminosities predicted during the first Gyr are around  
$3\times 10^{42}$ erg s$^{-1}$ \AA$^{-1}$.
This value is higher than the observed $L_{1500}$ of
$4.1\times 10^{41}h_{50}^{-1}$ erg s$^{-1}$ \AA$^{-1}$ found for \C2.
Note that the  $\propto (1+z)^{-4}$ dimming of the surface brightness 
with the redshift makes it difficult to detect 
the outer regions of  the \gal.
However, considering  the UV emission inside a projected radius
of 10 kpc, reduces only slightly the UV \lum\ ($L_{1500}(r<10\,{\rm kpc})
=2.5\times 10^{42}$ erg s$^{-1}$ \AA$^{-1}$).
On the other hand, a simple comparison between the SFR deduced
from the \hbeta\ line, assuming that the extinction at the \hbeta\ is
negligible, and the SFR deduced from the UV continuum,
indicates for \C2\ a correction factor for dust between 7 and 48 (P98).
These very high correction factors should not be taken at face value,
since
this simplistic approach furnishes some unphysical results, 
such as negative extinctions for some objects.
It would be interesting to consider
a dust extinction index based on the UV part of the spectrum,
the most easily accessible to observations of LBGs.
The effect of dust is to flatten the spectrum,
and the colour \gr\ provides a reliable measure of the UV slope
(at $z \approx 3$, the effective redshifts of the two filters, 
4740 and 6850 \AA, respectively, are translated to 1190 and 1710 \AA).
The comparison of the observed \gro\ colours to 
the \grc\ colours predicted by an unreddened continuous
\sfor\ model with absorption by the Lyman $\alpha$ forest,
allowed P98 to deduce dust correction factors
between $\sim1$ and $\sim 10$
for the UV luminosities of the LBGs in their sample.
In the case of \C2, 
a value  of  $L_{1500}=3.9(1.8) \times 10^{42}$ erg s$^{-1}$ \AA$^{-1}$ 
is obtained after a correction for dust extinction
assuming a Calzetti attenuation law and
a continuous $10^7(10^9)$ years old  \sfor.

In view of the importance of the \gr\ colour in checking for \sfor\ and
estimating the dust extinction, Figure 1 also shows  \grc\
predicted for \C2\  ($z=3.317$) by the fiducial model, obtained as
follows:
in the first place, the integrated SED is calculated for several
apertures,
using the Bruzual \& Charlot (1998) models; then the SED is redshifted to
$z=3.317$, reddened by the Lyman $\alpha$ forest opacity (Madau 1995), 
and convolved with the filter transmission curves. 
Finally, when \grc\ is bluer than the \gr\ colour of the \gal\
(\gro=1.13),
we calculate, assuming a Calzetti attenuation curve, the value of
$A_{1500}$
needed to match \gro. Since \gr\ becomes redder with time, we can use the
condition \grc$<$\gro\ to set an upper limit in the age of the \gal,
beyond which $A_{1500}$ becomes formally negative.
This limit is 1.00 Gyr, for an aperture $r<10$ kpc, and 1.52 Gyr,
if the aperture encompasses the whole \gal.
The predicted colours are bluer for the larger aperture because: 1)
metallicities
are typically $\sim 0.1$ solar for $r>10$ kpc, implying bluer colours for
the
star population; and 2) there is some \sfor\ in the outer parts of the
\gal\ as
the gas driven by the galactic wind is compressed on its way out of the
\gal.
At the peak of the SFR,  $A_{1500}$ reaches $\approx 2.15$,
within the range $A_{1500}=1.58-2.44$ deduced by P98 for a continuous
\sfor\
lasting from $10^9$ to $10^7$ yr.
Figure 1 also shows the observed value of $L_{1500}$ corrected for dust
extinction using the time-dependent value of  $A_{1500}$ obtained as
above,
and also the values corrected as in P98.
The agreement with the predictions of our models
both for the galaxy as a whole as for the inner 10 kpc is excellent.
Therefore, if our model \gal\ were at a redshift $\sim 3$,
it would be easily seen as an LBG.

In order to explore the recent availability of 
infrared imaging, tracing the rest-frame optical light,
Figure 2 shows the blue luminosity
of the fiducial model inside several projected radii.
The similarity of rest-frame optical and UV sizes,
indicated by the optical and near-infrared observations,
is reproduced by the predictions of our model:
the  half-light radii at the maximum of  SFR,
at 1500 \AA\ and in the blue band,
are 1.64 and 1.51 kpc, respectively.
It is useful,  due to the possibility of missing light from the
outer parts of the galaxy, to consider the half-light radii
with respect only to the inner 10 kpc of the \gal.
In this case, the  half-light radii at the SFR peak are
1.46 and 1.44  kpc, for 1500 \AA\ and  blue light, respectively.
Therefore, the optical morphology follows the UV morphology,
and the galaxy remains compact in the optical band.
Note, however, that the light {\it does not} trace the mass.
At the maximum of  SFR, the half-mass radius (=7.5 kpc) is much
larger than the half-light radius.
The \sfor\ does not follow the stellar mass, but instead it is
regulated by the gas flows (e.g., the \sfor\ within the inner kpc is fed
by the \cf\ towards the \gal\ centre).
The \sfor\ is not coordinated along the \gal: $\nu_{SF}$ in
the inner kpc reaches several $\times 10$ Gyr$^{-1}$, whereas
the $\nu_{SF}$ averaged over the whole \gal\ is slightly larger
than 1 Gyr$^{-1}$.
In view of this, estimating the mass from the half-light radius
will seriously underestimate the \gal\ mass.
P98 were suspicious of  having underestimating the mass
of \C2\ (the value they derive is $5.5\times 10^{10}$ \msun).
Here we quantify their suspicion, suggesting that the
mass underestimate could be a factor 4-5.
In fact, at the SFR peak, our model predicts 
not only half-light radii (whatever their definition)
that are very similar to the $1.7h_{50}^{-1}$ found for \C2,
but also a stellar velocity dispersion of 179 \kms, 
essentially  identical to the observed $\sigma_{em}=190 \pm 25$,
whereas the stellar mass of our \gal\ model is
$2\times 10^{11}$ \msun\ at this time.

 \begin{figure}
 \centerline{
 \psfig{figure=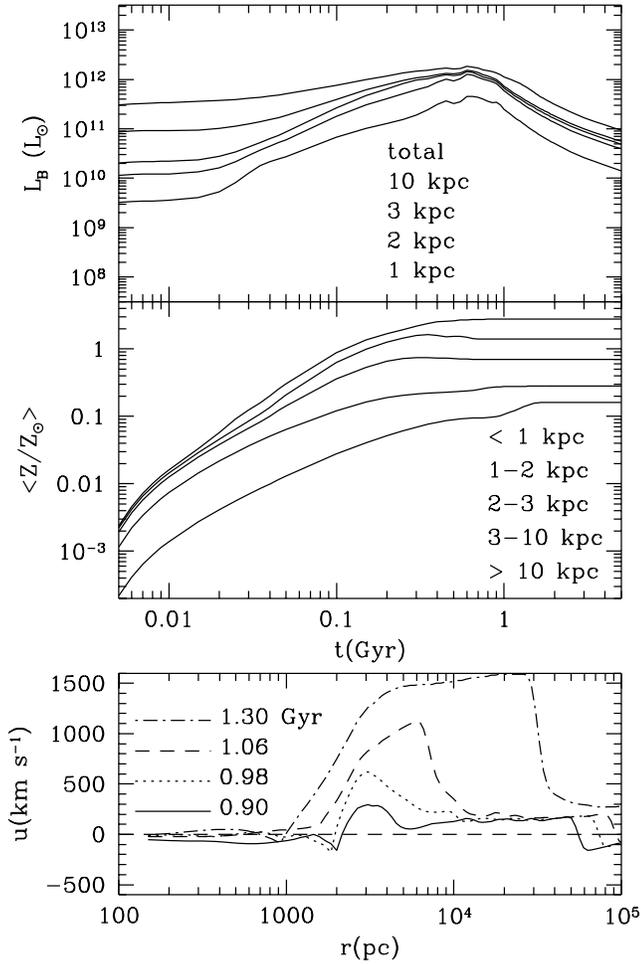,width=8.5cm,angle=0}}
 \caption{
Top panel: evolution of the blue luminosity of the fiducial
model inside several projected radii.
Middle panel: \ev\ of  the average metallicity of the stellar population,
within several spherical zones, illustrating the presence of metallicity
gradients
and the time scales for chemical enrichment.
The solar abundances are taken from Grevesse \& Anders (1989).
Lower panel:
\ev\ of the velocity profile, showing the onset of the outflow
at intermediate radii.
}
 \end{figure}

The metal lines in the spectra of LBGs, with origin in 
stellar photospheres,
interstellar absorption, and nebular emission,
indicate metallicities anywhere between 0.01 solar and solar
(Steidel et al. 1996).
On the other hand, the strong correlation between the UV spectral index
and metallicity in local starbursts would suggest a broad range in
metallicity 
from substantially subsolar to solar or higher (Heckman et al. 1998).
In order to make predictions on the metal content of LBGs,
Figure 2 also shows the average metallicity of the stellar population,
inside several spherical zones. The inner region reaches
solar metallicities (at $1.11\times 10^8$ and $1.56\times 10^8$
for the inner kpc and for the $1<r<2$ kpc region, respectively)
much earlier than the maximum in the SFR.
Therefore, when the \gal\ becomes visible as a LBG
(i.e. as a star-forming \gal), its metallicity inside a typical half-light
radius ($\sim 1.5$ kpc) will be solar or suprasolar.
On the other hand, substantial abundance gradients are built up.
The metallicity  approaches 3 \Zsun\ in the inner kpc,
while it is typically $\sim 0.1$ \Zsun\ for $r>10$ kpc.

Other important success of our models is the prediction
of important outflows during the stage of intense \sfor,
which could account for the outflow at a velocity of 
$500-1000$ \kms in the interstellar medium of  \C2\ suggested by 
the relative velocities
of the Lyman $\alpha$ emission lines and of the interstellar
absorption lines.
As a matter of fact, following the maximum of  the SFR,
an outflow appears at the intermediate radii,
between 2 and 10 kpc.
As we can see from Figure 2, once the outflow in the intermediate region
is established, outflow flows velocities of $500-1000$ \kms\ are
achieved for $t \sim 1$ Gyr.
After 1.17 Gyr, when the outflow reaches the \gal\ tidal radius
(i.e. the onset of the \gw), the wind velocity increases up to about 1900
\kms.
However, during the late \gw\ stage, the density in the outflowing gas
drops dramatically, making it difficult to obtain any signature
of the outflow via interstellar absorption lines and emission lines.
The flow structure is complex, because at inner radii
there is a highly subsonic (inflow velocity $\la 10$ \kms)
 \cf, and through the outer tidal there is
infall of low density gas proceeding at 60 \kms.
Therefore, 
the high density, high velocity 
outflowing gas in the intermediate region just after the peak in  the SFR
explains the large scale outflows with velocities of
$\approx 500$ \kms, deduced from the relative redshifts
of the interstellar absorption and Lyman $\alpha$ emission lines,
which are a common feature of LBGs (P98).

The success of our model for a $\sim L^*$ \sph\ or \el\ \gal\ in
reproducing several properties of  the LBG \C2\ gives additional support
to the
scenario in which LBGs are the progenitors of present-day bright
spheroids. 
High angular resolution spectroscopy will in the future provide
important information regarding the velocity field and angular momentum
of LBGs and help us to discern if they are young bulges or young
ellipticals.

\section{discussion}

The agreement
of the fiducial model with the properties of DSF2237-C2 
suggests that the mass range 
of LBGs does include present day $\sim L^*$ objects.
Note that the present model 
not only accounts for this particularly massive LBG
but also successfully predicts the properties
of the ensemble of the LBGs, within the scenario in which 
they are the progenitors of
the present day spheroids  with $0.1 L^* \la L_B \la L^*$.
Our models also reproduce the main properties 
of the four LBGs with lower $\sigma_{em}$'s in P98.
This is illustrated in Figure 1, in which the SFRs
deduced for these LBGs are similar to those of models
with  $M_G = 10^{10}-5\times 10^{10}$ \msun, for $t \la 1-1.5$ Gyr
(present day $-M_B=18.4-19.7$).

These models chosen because they exhibit
during the period $0.2 \leq t  \leq 1.5$ Gyr
(the lower limit on time garantees that a significant stellar component
has already been formed, and
for times later than the upper limit, THE SFR has probably decreased
below levels typical of LBGs)
the stellar velocity dispersion coincides with the 
values of $\sigma_{em}$ of the 4 low $\sigma_{em}$ LBGs of P98
(which are in the range $55 \pm 15 - 85 \pm 15$ \kms).

One of the central aspects of our modelling is that it follows in detail 
the impact of gas flows on the early \ev\ of the \gals.
Besides the importance of galactic winds in \gal\ \ev, as already
highlighted in the pioneering work of Larson (1974),
cooling flows also play a central role in \gal\ \ev\ ---
feeding a central AGN hosted in the galatic core,
building up metallicity gradients (FT),
and maintaining a moderate level of \sfor\ in the inner regions of the
\gal\ at late times, i.e. when the major stellar population of 
the \el\ \gal\ has already been formed (Jimenez et al. 1998).

As a matter of fact, the flow structure is complex,
exhibiting, for instance, during a considerable span of
the \gal\ \ev\ a partial wind, with inflow in the inner parts of the
\gal\ and outflow in the outskirts of the \gal.

Moreover, the flow structure varies with time, and the
same star-forming \gal\ can exhibit a variety of flow profiles,
depending on the evolutionary stage being picked up by the
observation.
As can be seen from Figure 2, 
in the fiducial model the outflow does not occur during the
whole period of intense \sfor. Outflow velocities of $\sim 500$ \kms\ are
achieved only $\sim 0.3$ Gyr after the maximum of the SFR and of
$\sim 1000$ \kms\ $\sim 0.4$ Gyr after the maximum.
The delay between the maximum of the SFR and the onset of the outflow
reflects the time needed for the energy input by SNe into the ISM
to overcome the gravitational binding energy of the gas.
It is possible the observation of LBGs, i.e. with a high SFRs,
both in the phase of outflow and before the onset of the outflow.
For earlier times, there are global
inflows, reaching velocities of up to a few 100 \kms.
Therefore, we expect a large dispersion in the relative redshifts
of the interstellar absorption, nebular emission
and Lyman $\alpha$ emission lines of LBGs.
As we discuss below, this seems to be the case.

 \begin{figure}
 \centerline{
 \psfig{figure=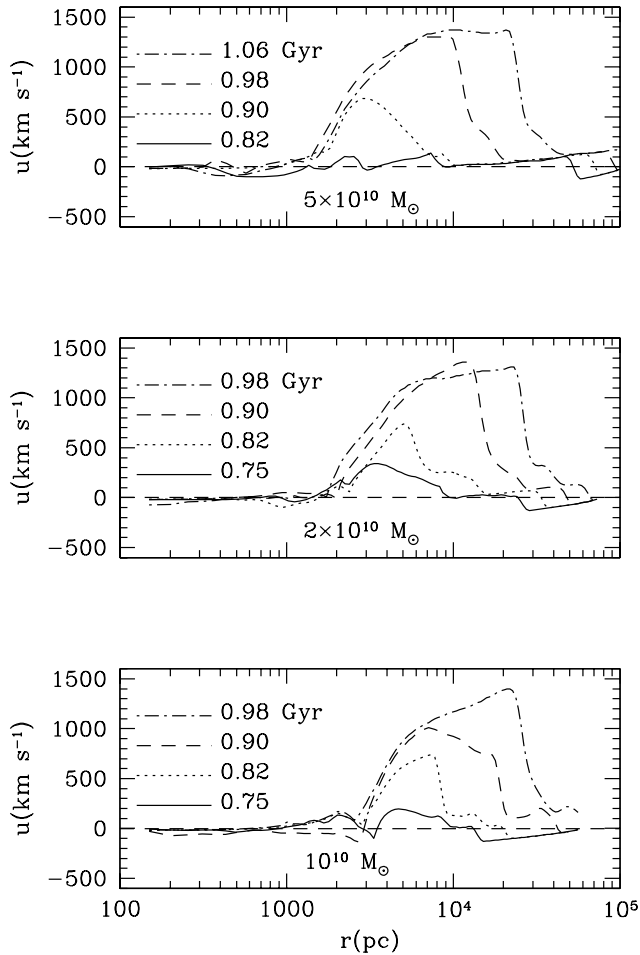,width=8.5cm,angle=0}}
 \caption{
Evolution of the gas velocity profiles
of low mass models aimed at describing the
$\sigma \approx 70$ \kms\ \gals\ of P98.
}
 \end{figure}

We can see from Figure 3 that for the models
with $M_G$ in the range $10^{10}-5\times 10^{10}$ \msun,
which describe well the four $\sigma_{em} \approx 70$ \kms\  in P98,  
the \ev\ of radial  flows is qualitatively similar
to that of the fiducial model.
The main difference is that the outflow happens earlier,
and, once the outflow is established, velocities higher
than 1000 \kms\ are reached faster.
This is a result of the shallower potential well of these \gals.
Note however, that the final wind velocities are somewhat lower than
in the fiducial model.

Assuming that the Balmer and [O~III] emission lines are 
at the galaxy systemic redshift,
the velocity shifts  of the interstellar absorption,
nebular emission, and \lya\ emission lines found by P98 indicate
that large velocity fields are a common feature of LBGs.
It is important to note that exactly the same result
has been found in nearby HII galaxies (Kunth etal 1998).
In all cases where \lya\ emission was  detected, 
the line peak is shifted by $\approx 1000$~\kms\ 
relative to the metal absorption lines. 
In two out of three cases
(Q0000$-$263 D6 and  B2~0902$+$343~C6) the nebular lines are at
intermediate 
velocities,
with Q0000$-$263 D6 exhibiting a conspicuous P-Cygni profile.
The most ready interpretation of these characteristics is the presence of
large 
scale outflows with velocities of $\sim 500$~\kms\ in the interstellar 
media of the galaxies observed:
the \lya\ emission is suppressed by 
resonant scattering and the only \lya\ photons escaping 
unabsorbed in our direction are those 
back-scattered from the far side of the expanding 
nebula, whereas in absorption against the stellar continuum
we see the approaching part of the outflow.
Within this scenario, the relative velocities of the three
line sets of these LBGs are consistent with our model, since
velocities of $\sim 500$~\kms\ are easily reached in 
the low mass models, at a galaxy age of $0.8-0.9$ Gyr. 

However, this simple symmetric picture probably does not account for 
all the variety of possible situations, since 
in the third LBG with \lya\ emission 
(the high $\sigma_{em}$ DSF~2237$+$116~C2) 
\hbeta\ and [O~III] emission are apparently at 
roughly the same velocity as the absorption lines, even though 
\lya\ emission is redshifted by $\approx 1000$~\kms. 
Also this case could be explained by our models with the 
\lya\ emission originating in the high velocity shell
receding in the back of the galaxy (in the fiducial model, at $t=1.1$ Gyr 
this shell is at $r\approx 10$ kpc with a velocity of 1250 \kms),
while the interstellar absorption lines could arise in the
few central kpc, with much lower expansion velocities. 
See the interesting discussion regarding escape of \lya\ photons
in HII galaxies by Kunth et al (1999).

In the P98 sample, 2 in 5 objects do not show \lya\ emission.
In one of these systems (Q0201+113 B13)
the interstellar absorption lines are {\it redshifted}  
by $250$ \kms\ with respect to
the \halpha\ emission (this object is at $z\approx 2.2$ and,
as a consequence, \halpha\ is observed instead of \hbeta).
If this velocity difference is real
(no error is quoted by P98 for the relative velocity of this 
object), within the
scenario depicted by our model, we could be observing the galaxy during 
its early global inflow.
The remaining object in P98 sample, also with no \lya\ emission,
Q0201$+$113 C6, shows a
$3200$\kms\ difference between emission and absorption lines,
too large to be accounted for any for our models. 
As a matter of fact, this difference is so large
that the line set identified as interstellar absorption
could be in reality an intervening absorption line system.
We would like to point that our predictions regarding  the importance
of the gas velocity field in the escape of \lya\ photons are similar to 
those of Kunth and collaborators for nearby HII galaxies 
(Kunth et al. 1998, 1999).

It is of interest to compare the results of the present model
with the predictions for LBGs of the semi-analytic models
of Baugh et al. (1998) and Mo, Mao, and White (1998)
which are based on disk formation models.
These models are simpler than the present model and, therefore,
their predictions are more limited.
For instance,  Mo et al. (1998) do not discuss gas flows.
However, in principle, their model could allow
for some global characterization of the infall associated
with the formation of a disk.  
Note that our prediction of outflows allows one
to distinguish the present model from disk models, since 
disk models include infall but do not exhibit outflows.

One success of
the models of Baugh et al. (1998) and Mo et al. (1998)
is the good description of
the clustering properties of the LBGs
within the framework of the most popular hierarchical models
for structure formation.
Note these success refers in fact to the clustering of halos,
independently if they host a spheroidal or a disklike
star-forming central \gal.
Mo et al. (1998) claims to explain also
the moderate SFRs, the small sizes
and the velocity dispersions of the LBGs.
However, as pointed out by Mo et al. themselves,
their calculation of the SFR is very sketchy.
In addition, their correct prediction of the compact size of the LBGs
is a consequence of identifying the LBGs with low angular momentum
objects, and so with small size for their mass.
In connection to this last point, one of the parameters of the
individual galaxies in Mo et al. (1998) is $\lambda$,
the spin parameter of the halo, and for systems with $\lambda$
smaller than some critical $\lambda_{crit}$, the gas cannot settle
into a centrifugally suppoted disk without first becoming
self-graviting.
The final configuration is probably spheroidal rather than disk-like.
In view of this, Mo et al. (1998) admits that a sizeable fraction
of LBGs could be spheiroids (see their Section 3.7), although
their model was initially designed for the formation of disks.
It is possible the existence of a population of disk objects among the
LBGs,
although the morphological studies of LBGs suggest that disks are
a minority in the LBG population.
{\it HST} imaging
(filters F606W to F814W, probing the rest frame UV range $1400-1900$ \AA)
has shown that the LBGs do not seem to have disk morphology,
with the exception of a few objects without central concentration,
for which an exponential profile provides a good fit to their surface
brightness distribution  (Giavalisco et al. 1996).
In addition, some objects have a light profile  following a $r^{1/4}$ law
over a large radial range.
Near-infrared {\it HST} NICMOS imaging 
(sensitive to the rest-frame optical light)
provides a similar picture (Giavalisco, private communication).
Future imaging observations would be very useful to establish
which fraction of the LBGs are disk-like systems.

With respect to the velocity dispersions of the LBGs,
Mo et al. (1998) reproduces the median values of the LBGS,
but runs into some problems with the $\sigma \approx$ 200 \kms\ of \C2.
since their predicted stellar velocity dispersions are typically around
70 \kms\ for their $\Omega_0=1$ cosmology and 120 \kms\ for
their $\Omega_0=0.3$ flat cosmology. 
In the distribution of velocity dispersions predicted by Mo et al. 
(their Figure 8), the highest velocity dispersion
is $\sim 170$ \kms\ for the $\Omega_0=1$ cosmology and, even for
the $\Omega_0=0.3$ flat cosmology (which predicts higher velocity
dispersions), 
values of $\sigma \approx 200$ \kms\ are very rare.
 
One interesting possibility offered by our modelling of the \gr\ colors
is obtaining constraints on the age of the LBGs within the scenario of
the young spheroid model. 
One minimal constaint is that
the galaxy could not be older than $\sim 1.5$ Gyr, otherwise \gr\ would
be redder than the observed. If, in addition to that, we require
that the SFRs predicted for our model be consistent with 
the SFR derived from the \hbeta\ emission, $SFR_{\hbeta}$,
the age of the LBG is constrained to be not older than $\sim 1$ Gyr.
In this case, the predicted values for $A_{1500}$,
$2.9-1.5$ (for $t$ from 0.1 to 1 Gyr),
are comparable to those
obtained by P98, because the \ev\ of the SFR of the fiducial model, 
with the constraint $t \la 1$ Gyr and
the strong decrease of the SFR beyond this time, includes
their simple models of continuous \sfor\ lasting for $10^7$ yrs and $10^9$ yr.
On the other hand, P98 estimated a value of $A_{1500}=4.2$, 
from comparing the SFRs deduced from \hbeta\ and UV luminosities, assuming
a Calzetti attenuation law and  continuous \sfor\ for $10^9$ yrs.
This value would represent a discrepancy with the values of $A_{1500}$
predicted by our models, were it not for the fact that the simple
comparison of \hbeta\ and UV luminosities could lead to unreliable
estimates of $A_{1500}$, as illustrated by some unphysical results,
e.g. negative extinctions found by P98 for some objects.
In determining $A_{1500}$ from the $SFR_{\hbeta}/SFR_{UV}$ ratio,
one should be aware of the uncertainties in converting $L{1500}$ to
SFR and that the long wavelength baseline involved in this estimation
increases the uncertainties of assuming one particular extinction law.

Finally,
the relatively high metal abundances obtained by our models,
could be expected with some hindsight, if we had considered the
high star formation rates derived for the LBGs, together with
the continued (for periods $\sim 1$ Gyr long) star formation 
favoured in previous works (e.g. Steidel et al. 1996).
One of the important consequences of these relativelly high metallicities
coupled to the large outflow velocities is that the halo of the
galaxy will be enriched in metals in $\sim 10^8$ yr, 
and the circungalactic environment out to distances of several hundreds
of kpc of the \gal\ will be contaminated in metals in $\sim 1$ Gyr or
less.
This fast chemical enrichment mechanism for the galactic halo
and in intergalactic medium could explain the chemical abundances 
of quasar absorption line systems, taken as a probe
of the gaseous galactic halo and of the intergalactic medium 
as for instance,
in Lyman limit systems (Viegas \& Fria\c ca 1995)
and in the \lya\ forest (Fria\c ca, Viegas \& Gruenwald 1998).

\section*{Acknowledgments}

We thank Gustavo Bruzual for making us available the GISSEL code
for evolutionary stellar population synthesis.
We thank Mauro Giavalisco for supplying us with the filter transmission
curves of the $U_n G {\cal R}$ system.
A.C.S.F. acknowledges support from the Brazilian agencies
FAPESP, CNPq, and FINEP/PRONEX.
We would like to thank the anonymous referee,
whose suggestions greatly improved this paper.

\bsp

\end{document}